\documentclass[conference]{IEEEtran}
\IEEEoverridecommandlockouts
\usepackage{cite}
\usepackage{amsmath,amssymb,amsfonts}
\usepackage{algorithmic}
\usepackage{graphicx}
\usepackage{subfigure}
\usepackage{textcomp}
\usepackage{xcolor}
\usepackage{bm}
\usepackage{float}

\usepackage{multirow}
\usepackage{threeparttable}
\usepackage{breqn}
\usepackage{tikz}

\def\BibTeX{{\rm B\kern-.05em{\sc i\kern-.025em b}\kern-.08em
    T\kern-.1667em\lower.7ex\hbox{E}\kern-.125emX}}

\def\BibTeX{{\rm B\kern-.05em{\sc i\kern-.025em b}\kern-.08em
    T\kern-.1667em\lower.7ex\hbox{E}\kern-.125emX}}

\begin{document}

\IEEEoverridecommandlockouts
\IEEEpubid{\makebox[\columnwidth]{979-8-3315-4127-9/24/\$31.00~\copyright2024 IEEE \hfill} \hspace{\columnsep}\makebox[\columnwidth]{ }}

\title{Mixed Delay/Nondelay Embeddings Based Neuromorphic Computing with \\ Patterned Nanomagnet Arrays}

\IEEEpubidadjcol

\author{\IEEEauthorblockN{Changpeng Ti\IEEEauthorrefmark{1}, Usman Hassan\IEEEauthorrefmark{1}, Sairam Sri Vatsavai\IEEEauthorrefmark{1}, Margaret McCarter\IEEEauthorrefmark{1}\IEEEauthorrefmark{2}, Aastha Vasdev\IEEEauthorrefmark{1}\IEEEauthorrefmark{3}, Jincheng An\IEEEauthorrefmark{4} \\ Barat Achinuq\IEEEauthorrefmark{2}, Ulrich Welp\IEEEauthorrefmark{3}, Sen-Ching Cheung\IEEEauthorrefmark{1}, Ishan G Thakkar\IEEEauthorrefmark{1}, J. Todd Hastings\IEEEauthorrefmark{1}}
\IEEEauthorblockA{\IEEEauthorrefmark{1}Department of Electrical and Computer Engineering, University of Kentucky, Lexington, KY 40506, USA}
\IEEEauthorblockA{\IEEEauthorrefmark{2}Advanced Light Source, Lawrence Berkeley National Laboratory, Berkeley, CA 94720, USA}
\IEEEauthorblockA{\IEEEauthorrefmark{3}Materials Science Division, Argonne National Laboratory, Lemont, IL 60439, USA}
\IEEEauthorblockA{\IEEEauthorrefmark{4}Department of Physics, University of Kentucky, Lexington, KY 40506, USA}

\IEEEauthorblockA{\textit{
cti222@uky.edu, usman.hassan@uky.edu, igthakkar@uky.edu, todd.hastings@uky.edu}}}


\maketitle

\begin{abstract}
Patterned nanomagnet arrays (PNAs) have been shown to exhibit a strong geometrically frustrated dipole interaction. Some PNAs have also shown emergent domain wall dynamics. Previous works have demonstrated methods to physically probe these magnetization dynamics of PNAs to realize neuromorphic reservoir systems that exhibit chaotic dynamical behavior and high-dimensional nonlinearity. These PNA reservoir systems from prior works leverage echo state properties and linear/nonlinear short-term memory of component reservoir nodes to map and preserve the dynamical information of the input time-series data into nondelay spatial embeddings. Such mappings enable these PNA reservoir systems to imitate and predict/forecast the input time series data. However, these prior PNA reservoir systems are based solely on the nondelay spatial embeddings obtained at component reservoir nodes. As a result, they require a massive number of component reservoir nodes, or a very large spatial embedding (i.e., high-dimensional spatial embedding) per reservoir node, or both, to achieve acceptable imitation and prediction accuracy. These requirements reduce the practical feasibility of such PNA reservoir systems. To address this shortcoming, we present a mixed delay/nondelay embeddings-based PNA reservoir system. Our system uses a single PNA reservoir node with the ability to obtain a mixture of delay/nondelay embeddings of the dynamical information of the time-series data applied at the input of a single PNA reservoir node. Our analysis shows that when these mixed delay/nondelay embeddings are used to train a perceptron at the output layer, our reservoir system outperforms existing PNA-based reservoir systems for the imitation of NARMA 2, NARMA 5, NARMA 7, and NARMA 10 time series data, and for the short-term and long-term prediction of the Mackey Glass time series data.

\end{abstract}


\section{Introduction}
Patterned nanomagnet arrays (PNAs) are typically fabricated by etching nanomagnets in ferromagnetic thin films (e.g., permalloy films) in the desired geometric arrangement using lithography \cite{schiffer_artificial_2021}. Each nanomagnet in a PNA behaves as a microscopic Ising spin whose dipole interaction with the neighboring spins is geometrically frustrated \cite{hu_distinguishing_2023}. Such frustrated interactions often lead to numerous nearly degenerated energy states in the PNA because of which not all spins in the PNA attain the ground-energy state when the PNA is subjected to an external magnetic field perturbation \cite{skjaervo_advances_2020}. The lingering of nanomagnet spins in the degenerated energy states gives rise to richly nonlinear magnetization dynamics in PNAs \cite{skjaervo_advances_2020}. 

The nonlinear magnetization dynamics give rise to the echo-state property and nonlinear/linear short-term memory in PNAs. These properties enable PNAs to memorize and adaptively learn spatio-temporal features and hidden patterns in complex time series data when the time series data are encoded and applied to PNAs as a sequence of perturbing external magnetic fields \cite{skjaervo_advances_2020,stenning_adaptive_2023}. These capabilities of PNAs make them very attractive for realizing neuromorphic reservoir systems that can be used to imitate and predict time series data \cite{allwood_perspective_2023,taniguchi_spintronic_2022,ustinov_current-controlled_2024,vidamour_reconfigurable_2023,lee_task-adaptive_2024,kobayashi_thermally-robust_2023,hu_distinguishing_2023,edwards_passive_2023,stenning_neuromorphic_2023-2,gartside_reconfigurable_2022-2,jiang_task-adaptive_2024}. 

Designing a quality reservoir system using PNAs requires probing the magnetization dynamics and the nearly degenerated spin energy states of PNAs so that the spatial embeddings of the dynamical information of the input time series data, preserved in the magnetization states of PNAs, can be extracted for postprocessing \cite{hu_distinguishing_2023,yan_emerging_2024}. Such spatial embeddings are referred to as nondelay spatial embeddings as they provide the preserved dynamical information at a particular time step of the input time-series data \cite{hou_harvesting_2022,duan_embedding_2023}. To extract nondelay spatial embeddings of PNA-based reservoir systems, the orientations of nanomagnet spins in PNAs can be probed by direct magnetic imaging of PNAs \cite{skjaervo_advances_2020}. In addition to magnetic imaging, other common tools for probing PNAs and extracting nondelay spatial embeddings include ferromagnetic resonance \cite{stenning_adaptive_2023,gartside_reconfigurable_2022-2,jiang_task-adaptive_2024}, resonant soft X-ray scattering, X-ray photon correlation spectroscopy, and magnetoresistance transport measurement (using giant magnetoresistance (GMR) or anisotropic magnetoresistance (AMR) or tunneling magnetoresistance (TMR)) \cite{hu_distinguishing_2023}. Several prior works have used these tools to demonstrate PNA-based reservoir computing capabilities \cite{allwood_perspective_2023,taniguchi_spintronic_2022,ustinov_current-controlled_2024,vidamour_reconfigurable_2023,lee_task-adaptive_2024,kobayashi_thermally-robust_2023,hu_distinguishing_2023,edwards_passive_2023,stenning_neuromorphic_2023-2,gartside_reconfigurable_2022-2,jiang_task-adaptive_2024}.


Unfortunately, however, these tools face several shortcomings that diminish their use in practical reservoir computing systems and applications. For instance, the ferromagnetic resonance and X-ray based tools are powerful for (i) distinguishing the nearly degenerated spin energy states, (ii) understanding the physics of PNAs, and (iii) extracting the nondelay spatial embeddings of reservoir states that are rich in the spatially separable dynamical information of the input data, but they are not practical for co-packaged hardware-based implementations of reservoir computing applications. On the other hand, the TMR/GMR/AMR-based transport measurements are amenable to hardware-based implementations of practical reservoir computing applications, but they provide nondelay spatial embeddings of reservoir states that are scarce in the spatially separable dynamical information of input data. This scarcity of spatially separable dynamical information results in poor accuracy for imitation and prediction tasks. 

To address these shortcomings, we propose a mixed delay/nondelay spatial embeddings-based PNA reservoir system. Our system extracts the nondelay spatial embeddings of PNA reservoir states using TMR-based probing of PNA spins and augments the extracted nondelay spatial embeddings with delayed spatial embeddings from a total of past h=50 time steps. The combination of delay and nondelay-based spatial embeddings dramatically increases the richness of the spatially separable dynamical information readily available to our reservoir system, increasing the imitation and prediction accuracy of our system compared to the TMR/GMR/AMR-based PNA reservoir systems from prior works. In addition, the TMR-based probing increases the practicality of our reservoir system compared to the ferromagnetic resonance and X-ray probing-based PNA reservoir systems from prior works. 

This paper makes the following key contributions. 

\begin{itemize}
    \item We propose a TMR probing-based PNA reservoir system that combines delay and nondelay spatial embeddings to improve the practicality and accuracy of reservoir computing;
    \item We evaluate our proposed PNA reservoir system for imitation of the NARMA 2, NARMA 5, NARMA 7, and NARMA 10 time series data;
    \item We evaluate our proposed PNA reservoir system for short-term and long-term prediction of the Mackey-Glass time series data;
    \item We explore the accuracy impacts of using a single-layer versus multilayer perceptron at the output layer of our PNA reservoir system;
    \item We compare the imitation and precision accuracy results for our system with the results for five different reservoir systems from prior works. 
\end{itemize}

\section{Preliminaries}

\subsection{Neuromorphic Reservoir Computing: Overview}
\begin{figure}
    \centering
    \subfigure[]
    {
        \includegraphics[width=\linewidth]{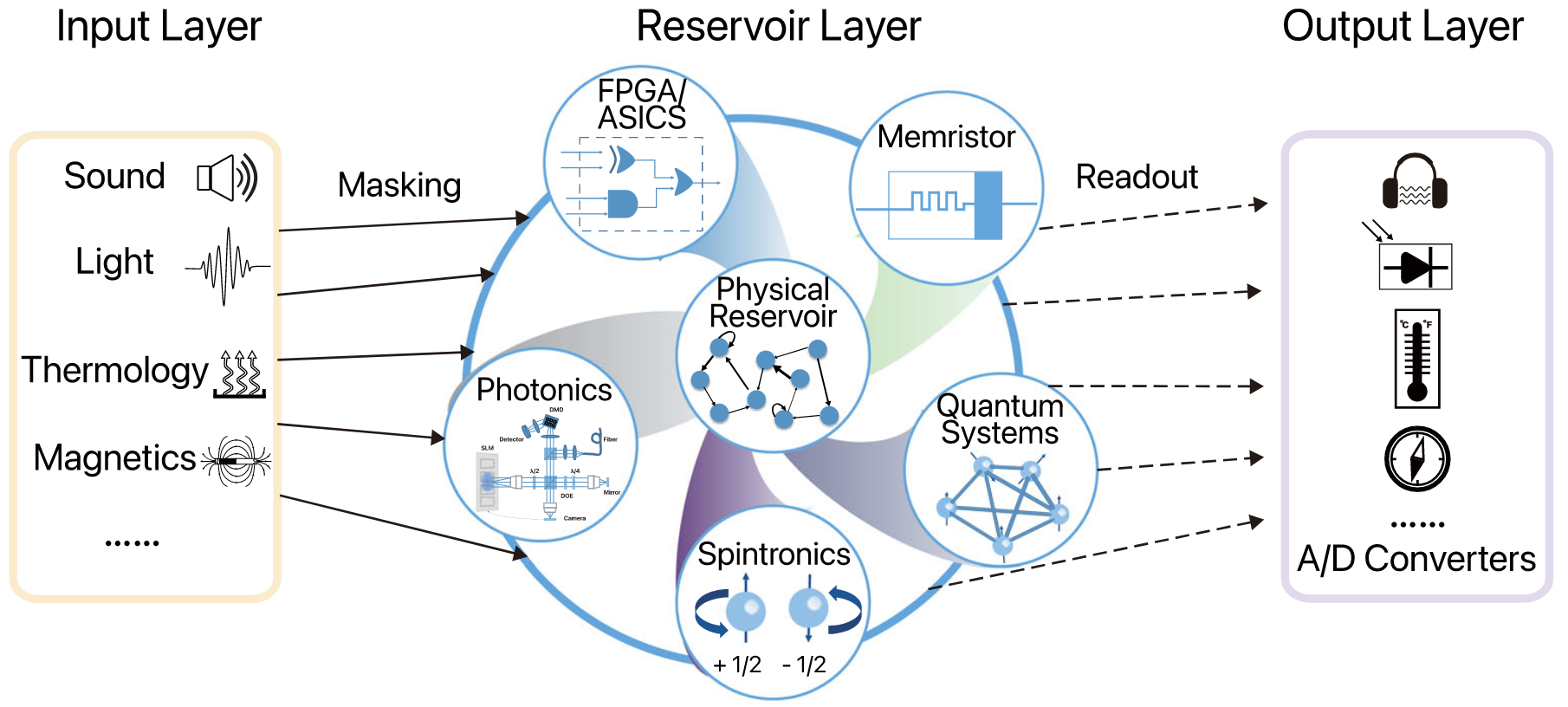}
        \label{fig:RC-concept}
    }
    \subfigure[]
    {
        \includegraphics[width=\linewidth]{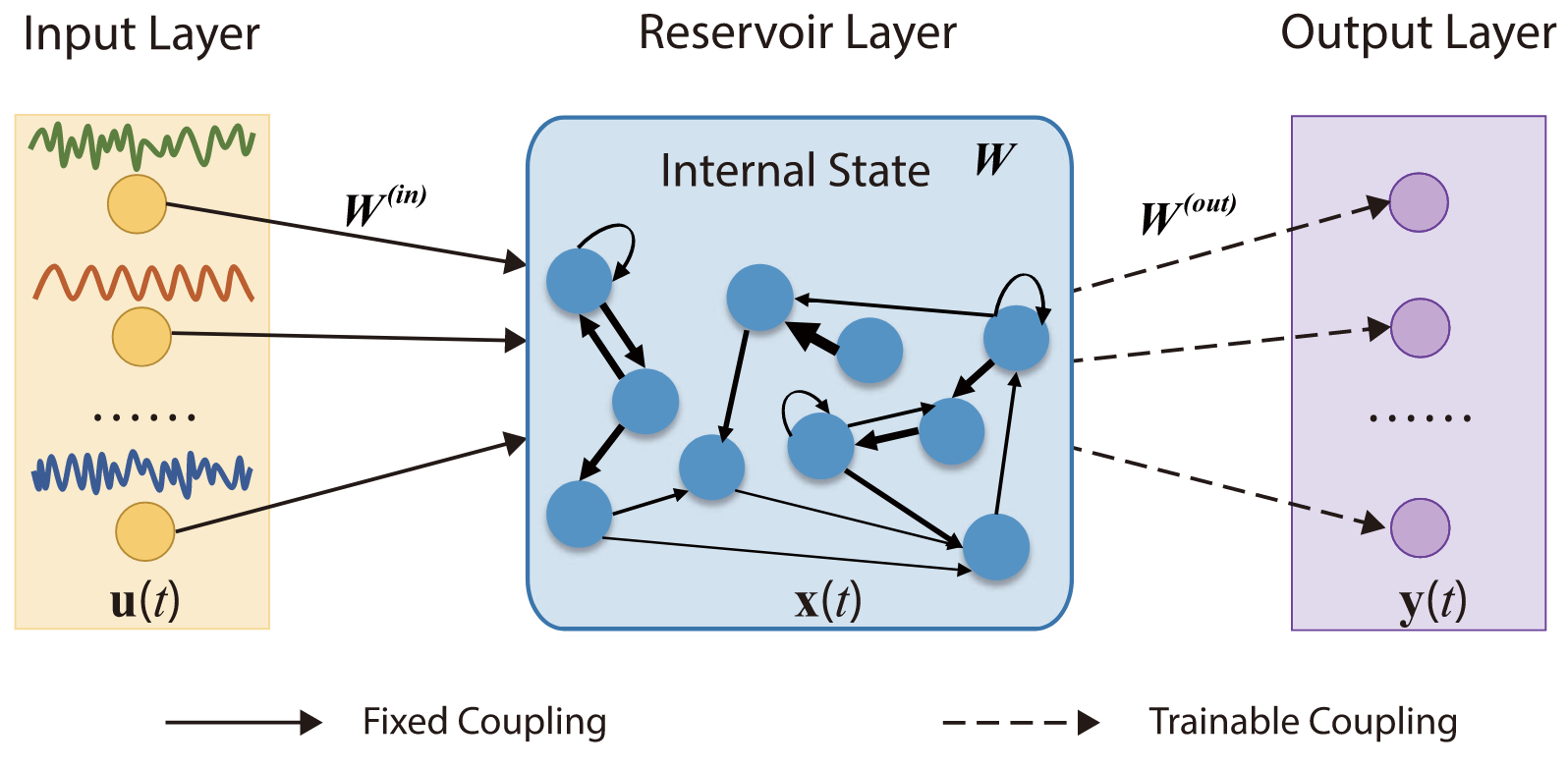}
        \label{fig:RC-flow}
    }
    \caption{(a) A schematic representation of the core idea of physical reservoir computing. (b) A schematic illustration of the computational flow of reservoir computing. (a) and (b) are reproduced from \cite{yan_emerging_2024}.}
    \label{fig_RC-schematic}
\end{figure}

The core idea of the neuromorphic reservoir computing is to use a neuromorphic dynamical system as a physical reservoir to adaptively learn spatio-temporal features and hidden patterns in complex time series \cite{yan_emerging_2024,vatsavai_silicon_2021} (see Fig. \ref{fig:RC-concept}). A perspective on emerging challenges and opportunities for reservoir computing is available in \cite{yan_emerging_2024}. As described in \cite{yan_emerging_2024}, the basic process of reservoir computing involves preprocessing the external input signal, mapping the input signal onto interacting physical nodes within the reservoir to transform it into complex spatiotemporal patterns in a high-dimensional space, probing the physical reservoir state, and finally postprocessing the reservoir state to generate output data. Various physical devices and systems with disparate physical phenomena can be used to realize a reservoir \cite{vatsavai_silicon_2021,chen_emerging_2023,fang_-materio_2023,gauthier_next_2021-3,yan_emerging_2024} (see Fig. \ref{fig:RC-concept}).

The computational flow of reservoir computing comprises three layers, namely the input, reservoir, and output layers (see Fig. \ref{fig:RC-flow}). At the input layer, the raw inputs are encoded into temporal sequences of physical perturbations that can directly excite the dynamics of the nonlinear physical nodes of the reservoir. The complex interaction dynamics of the physical nodes of the reservoir, governed by predefined physical rules, determine the reservoir state. This reservoir state is then transformed into the reservoir output using a perceptron typically a single-layer linear perceptron. In a standard setup, the input encodings and the physical rules defining the dynamical behavior of the reservoir are fixed, and only the perceptron at the output layer is trained.

To achieve high accuracy from a reservoir system, it is generally desirable to have a very high- dimensional reservoir state. The dimensionality of the reservoir state referes to the number of embeddings of spatio-temporally dynamical information in the reservoir state vector. Typically, there is a one-to-one correspondence between the number of embeddings in the reservoir state vector and the number of physical nodes in the reservoir system; therefore, increasing the dimensionality of the reservoir state often results in a large number of physical nodes in the reservoir, making it less feasible to implement. To make reservoir systems more practical and encourage their wider adoption, it is essential to increase the dimensionality of the reservoir state without proportionally increasing the number of physical nodes. Fortunately, previous studies \cite{hou_harvesting_2022,duan_embedding_2023,ma_randomly_2018-1,rand_detecting_1981,koster_insight_2021-1} have demonstrated that this can be achieved by introducing time-delayed embeddings in addition to the general nondelay embeddings per physical reservoir node. \textit{In this paper, we investigate the use of time-delayed embeddings in PNA-based reservoir systems for the first time}.

\subsection{Computing with Patterned Nanomagnet Arrays (PNAs)}

PNAs can be manufactured using standard lithography techniques in various geometric topologies and structures (e.g., square \cite{jensen_flatspin_2022}, pinwheel \cite{jensen_reservoir_2020}, kagome \cite{frotanpour_vertex_2021}topologies and various other structures \cite{skjaervo_advances_2020-1,heyderman_spin_2022,arnalds_new_2016-1,talapatra_coupled_2021-1,woods_switchable_2021-2,ribeiro_realization_2017,schiffer_artificial_2021}). Realizing any arbitrary more exotic PNA topologies is also possible, e.g., artificial quasicrystals \cite{bhat_ferromagnetic_2014,farmer_direct_2016}, distorted topologies \cite{frotanpour_angular-dependent_2021,frotanpour_magnetization_2020}, and connected nanomagnet networks where domain walls can travel through the network \cite{vidamour_reconfigurable_2023}. Individual nanomagnets of a PNA system can be realized in various shapes, e.g., elongated \cite{jensen_reservoir_2020,frotanpour_vertex_2021} and circular \cite{edwards_passive_2023}. 

\begin{figure}
    \centering
    \includegraphics[scale=0.65]{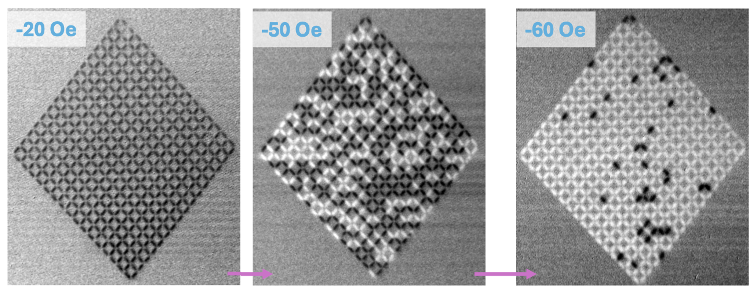}
    \caption{X-ray magnetic circular dichroism (XMCD) photoemission electron microscopy (PEEM) image of a square topology PNA under the influence of external magnetic field. Dark and light-colored nanomagnets in the PNA represent two spin directions of the nanomagnet dipoles. When the external magnetic field amplitude changes from -20 Oe to -50 Oe to -80 Oe, the spin directions of different numbers of nanomagnets flip due to the collective dynamical behavior of the nanomagnets.}
    \label{fig:pna_fab}
\end{figure}

In a PNA, the nanomagnets are typically arranged in a way that not all dipole interactions among the nanomagnets can be satisfied simultaneously. This gives rise to collective dynamical magnetization behavior, which can be driven by an external magnetic field. In Fig. \ref{fig:pna_fab}, we demonstrate an experimentally probed trigger of such dynamical magnetization behavior in an electron-beam lithography-fabricated sample of a square PNA. 

The collective dynamical behavior of the nanomagnets gives rise to a variety of emergent phenomena including magnetic monopoles, vertex-based frustration, phase transition, and chiral dynamics \cite{jungfleisch_dynamic_2016,frotanpour_angular-dependent_2021-5,lendinez_emergent_2021,lendinez_nonlinear_2022-2,kaffash_control_2020,bang_spin_2022,kaffash_direct_2022}. This collective dynamical behavior of PNAs has been shown to give rise to the echo state property and nonlinear/linear memory capacity in the PNAs. Due to these properties, PNAs have been used as high-quality neuromorphic reservoirs in several prior works \cite{allwood_perspective_2023,taniguchi_spintronic_2022,ustinov_current-controlled_2024,vidamour_reconfigurable_2023,lee_task-adaptive_2024,kobayashi_thermally-robust_2023,hu_distinguishing_2023,edwards_passive_2023,stenning_neuromorphic_2023-2,gartside_reconfigurable_2022-2}.

\subsection{Input Encoding and Readout for PNA Reservoirs}
Spin-wave spectral fingerprinting has been a common method to investigate the physics and computing capabilities of PNAs \cite{vanstone_spectral_2022,lee_task-adaptive_2024,gartside_reconfigurable_2022-2}. This is typically enabled by flip-chip integration of the PNA chip with an RF coplanar waveguide \cite{vanstone_spectral_2022}. The coplanar waveguide can be connected to a microwave generator to enable the coupling of an external RF magnetic field with the PNA sample, constituting a method for applying the input signal to the PNA-based reservoir \cite{gartside_reconfigurable_2022-2}. The coplanar waveguide can also be used to probe the ferromagnetic resonance spectra of the PNA sample, constituting a method for reservoir state readout. Apart from applying an external magnetic field via the coplanar waveguide, the input signal can also be applied through spin-orbit torque (SOT) or magnetic tunnel junction (MTJ)-based switching of the actuating nanomagnets in the PNA \cite{edwards_passive_2023,taniguchi_spintronic_2022}. Similarly, alternative methods also exist for the state readout of PNA-based reservoirs. These include magnetization probing using TMR \cite{edwards_passive_2023} or GMR/AMR \cite{hu_distinguishing_2023}. An analysis of the energy consumption of the SOT-based input application and the MTJ-based readout enabled by TMR has been carried out in \cite{edwards_passive_2023}, and evidently, the TMR-based readout and SOT-based input application can be substantially more energy efficient than the coplanar waveguide-based input and readout methods.

\section{Proposed Delay/Nondelay Embeddings Based PNA Reservoir System}


    

\subsection{Design}
\begin{figure}
    \centering
    \subfigure[]
    {
        \includegraphics[scale=0.465,page=1,trim={255 170 180 125},clip]{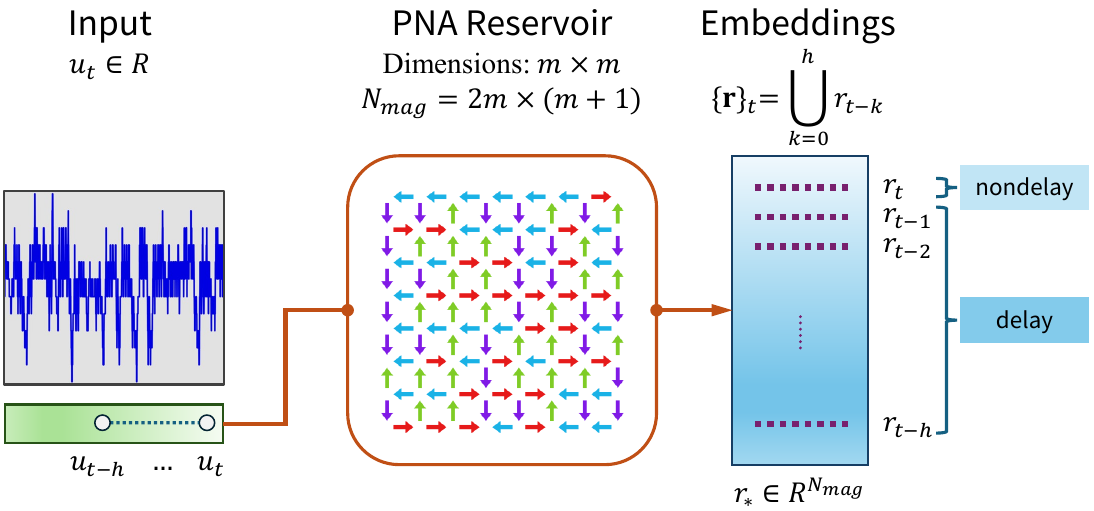}
        \label{fig_proposed:1a}
    }
    \subfigure[]
    {
        \includegraphics[scale=0.465,page=2,trim={245 155 255 150},clip]{figures_proposed.pdf}
        \label{fig_proposed:1b}
    }
    \caption{Schematic of our proposed delay/nondelay embeddings based PNA reservoir system. (a) Mixed delay/nondelay embeddings generated by our proposed PNA reservoir system. The embeddings are a combination of nondelay and delay reservoir responses. (b) The layout of the perceptron at the output of the system that enables time series imitation and prediction tasks.
    A multiplayer perceptron with SELU activation function is shown but our design also works well with a single linear layer perceptron.}
    \label{fig_proposed:1}
\end{figure}

Our proposed PNA reservoir system, illustrated in Fig. \ref{fig_proposed:1}, takes as input the time-series data $\{u_t\}_{t\in \mathbb{N}}$ (a sequence of sampled signal amplitudes). At each time step $t$, $u_t$ is encoded into $N$ cycles of global external magnetic field amplitude pairs, $(\Vec{h}_t^{ext}, -\Vec{h}_t^{ext})$. The magnetic field amplitudes of the field cycles corresponding to $u_t$ are linearly proportional to the amplitude of $u_t$. This method of encoding $u_t$ as cycles of magnetic field amplitudes is inspired from \cite{gartside_reconfigurable_2022-2,stenning_neuromorphic_2023-2}. We set N=5 for our PNA reservoir system, though other values may also be valid. Determining the optimal value of N is beyond the scope of this work. 

Our system consists of a single PNA reservoir node with configurable parameters such as geometric topology, dimensions $m \times m$, and magnetic properties. We have used square and pinwheel topologies \cite{jensen_flatspin_2022} in this work, though other geometric topologies could also be explored (e.g., kagome \cite{frotanpour_magnetization_2020-3,frotanpour_vertex_2021}). When the input-encoded magnetic field cycles are applied sequentially to the PNA reservoir node, they induce hysteretic and dynamical magnetization interactions among the constituent nanomagnets of the PNA. After the magnetization interactions in the PNA node stabilize, the magnetization status at each nanomagnet is probed. The physical probing of the magnetization of each nanomagnet can be achieved by using the Tunneling Magnetoresistance (TMR) effect \cite{edwards_passive_2023} or the Giant/Anisotropic Magnetoresistance (GMR/AMR) effects \cite{hu_distinguishing_2023}.
The magnetization values of all the nanomagnets in the PNA node are then combined into the reservoir node’s state vector, denoted as the nondelay embedding $r_t \in \Re^{2m \times (m+1)}$. This nondelay embedding represents the response of the reservoir system at the current time step. This embedding preserves the dynamical information of multiple past time steps of sequence $\{u_{t'}\}_{t'\leq t}$.
This preservation of dynamical information is due to the echo state property and linear/nonlinear short-term memory capacity of the PNA reservoir node, as expected.


In addition to this nondelay embedding, our system incorporates a delay embedding to explicitly capture historical information on the input signal dynamics for imitation and prediction. This delay embedding at time $t$ is formed by concatenating the reservoir's nondelay embeddings from past $h$ time steps, denoted as $r_{t-k}$ where $k = 1, 2, \dots, h$.
The final reservoir state at time $t$
 is the combination of delay and nondelay embeddings, represented by the concatenated state vector
 $\{\mathbf{r\}_t} = [r_t, r_{t-1}, \ldots, r_{t-h}]$.
 This concatenated state vector spatially encapsulates the dynamical behavior of the system up to the current time step, and is used for subsequent tasks such as signal imitation and prediction.

For the training process, a perceptron (either single-layer or multilayer) is used to transform the augmented state vector $\{\mathbf{r\}_t}$ into a target signal (sequence of sampled signal amplitudes). In the case of the imitation task, the output is trained to reproduce the target signal $\{y_t\}$. For prediction tasks, the output forecasts a time-advanced input signal $\{u_{t+\tau}\}$, where $\tau$ is the prediction horizon. Designating the target output signal (sequence) as $Y$ and the output signal (sequence) generated by the PNA reservoir system as $\bar Y$, we train the perceptron employing a linear combination of mean square error (MSE) and inverted Pearson correlation coefficient (CC) as the loss function:
\begin{equation}
L = \alpha \cdot \text{MSE}(Y, \Bar{Y}) + (1 - \alpha) \left[ 1 - \text{CC}(Y, \Bar{Y}) \right]
\label{eq:loss_function}
\end{equation}
Here, we take $\alpha$ = 0.15. Our approach allows the system to effectively capture temporal dependencies in the input signal, thereby enhancing its ability to imitate and predict time-series data with high accuracy.

\subsection{Simulation and Implementation}

For the implementation and evaluation of our PNA reservoir system, we model a PNA with size $m \times m = 7 \times 7$ and square topology (containing a total of $2m \times (m+1) = 112$ nanomagnets) using flatspin \cite{jensen_flatspin_2022}.
The size of the PNA represents a hyperparameter for controlling the reservoir quality. It directly influences the expressivity and computational complexity of the reservoir \cite{love_task_2021}. Increasing the PNA size is theoretically favorable for higher expressivity; however, in practice, realizing high-quality reservoirs (i.e., with high echo state property, memory capacity, and stability) with large PNAs requires exploration and co-optimization of several physical properties of PNAs, including the
shape, size,
and interspacing of the individual nanomagnets. We leave such exploration and co-optimization for future work and select the PNA size of $7 \times 7$ for this work because this size has been accepted as one of the standard PNA sizes in prior work \cite{jensen_flatspin_2022}.
We simulate the magnetization dynamics of this PNA from our flatspin-based model \cite{jensen_flatspin_2022} to extract the delay and nondelay spatial embeddings. We incorporate this flatspin-based PNA modeling environment with a PyTorch-based framework for input data pre-processing, perceptron training, testing, data analysis, and accuracy evaluation.
Flatspin \cite{jensen_flatspin_2022} does not
impose
a hard limit on the PNA size; however, its main drawback is the
rapidly increasing simulation time as the array size grows.
In the future, we plan to explore more efficient simulation platforms such as HotSpice~\cite{maes_design_2024,maes_hotspice_2024} in studying the effect of scaling the PNA size to thousands of nanomagnets. 

\section{Time Series Imitation}
\subsection{Method}
In the Non-linear Autoregressive Moving Average (NARMA) imitation task, we aim to replicate the behavior of NARMA $N$ where $N=(2, 5, 7, 10)$, where higher values of N represent systems with increased temporal complexity and nonlinear dependencies. The sequences $y_t$ are generated using the nonlinear difference equation from \cite{suzuki2022natural} with the same parameters: $\alpha=0.3, \beta=0.05, \gamma=1.5,$ and $\delta=0.1$. The input signal $u(t)$ is randomly generated by uniformly sampling between 0 and 0.5. Sequences generated from NARMA systems exhibit a noisy, cyclical behavior where increasing the order leads to more complex dependencies on past values. We set the history length to $h=50$, meaning that the mixed delay/nondelay embeddings include the current reservoir state as well as the reservoir states from the previous 50 time steps. The perceptron used for training consists of a single hidden layer with a SeLU activation function.

The goal is to estimate the target signal $y_t$ where $y_t$ is generated according to the NARMA equation, using the mixed delay/nondelay embedding $\{\mathbf{r\}_t}$ derived from the PNA reservoir system. The training is performed by minimizing a loss function combining the mean squared error (MSE) and the inverted Pearson correlation coefficient, as described in Eq. \ref{eq:loss_function}. We use data sequences of length $5000$, with $20\%$ of the input-output pairs randomly selected for testing and the remaining for training.

\begin{table}[H]
\centering
\caption{Prediction results for NARMA sequences of different orders (N) using two $7 \times 7$ topologies.}
\label{tab:NARMA_combined}
\begin{tabular}{|l|llllll|}
\hline
\textbf{N}      & \multicolumn{1}{l|}{\textbf{MSE}} & \multicolumn{1}{l|}{\textbf{NRMSE}} & \multicolumn{1}{l|}{\textbf{CC}} & \multicolumn{1}{l|}{\textbf{MSE}} & \multicolumn{1}{l|}{\textbf{NRMSE}} & \textbf{CC} \\ \hline
\textbf{}       & \multicolumn{3}{c|}{\textbf{Train (Square ICE)}}                                                              & \multicolumn{3}{c|}{\textbf{Test (Square ICE)}}                                         \\ \hline
2               & \multicolumn{1}{l|}{8.12e-2}      & \multicolumn{1}{l|}{3.67e-1}        & \multicolumn{1}{l|}{0.99905}     & \multicolumn{1}{l|}{1.63e-2}      & \multicolumn{1}{l|}{1.78e-1}        & 0.99956     \\ \hline
5               & \multicolumn{1}{l|}{7.44e-2}      & \multicolumn{1}{l|}{3.56e-1}        & \multicolumn{1}{l|}{0.99876}     & \multicolumn{1}{l|}{1.81e-2}      & \multicolumn{1}{l|}{1.87e-1}        & 0.99775     \\ \hline
7               & \multicolumn{1}{l|}{6.98e-2}      & \multicolumn{1}{l|}{3.73e-1}        & \multicolumn{1}{l|}{0.99892}     & \multicolumn{1}{l|}{2.36e-2}      & \multicolumn{1}{l|}{2.04e-1}        & 0.99733     \\ \hline
10              & \multicolumn{1}{l|}{6.44e-2}      & \multicolumn{1}{l|}{3.29e-1}        & \multicolumn{1}{l|}{0.99698}     & \multicolumn{1}{l|}{1.71e-2}      & \multicolumn{1}{l|}{1.69e-1}        & 0.99214     \\ \hline
\textbf{}       & \multicolumn{3}{c|}{\textbf{Train (Pinwheel ICE)}}                                                          & \multicolumn{3}{c|}{\textbf{Test (Pinwheel ICE)}}                                      \\ \hline
2               & \multicolumn{1}{l|}{8.01e-2}      & \multicolumn{1}{l|}{3.64e-1}        & \multicolumn{1}{l|}{0.99957}     & \multicolumn{1}{l|}{1.57e-2}      & \multicolumn{1}{l|}{1.75e-1}        & 0.99905     \\ \hline
5               & \multicolumn{1}{l|}{7.31e-2}      & \multicolumn{1}{l|}{3.53e-1}        & \multicolumn{1}{l|}{0.99918}     & \multicolumn{1}{l|}{2.21e-2}      & \multicolumn{1}{l|}{2.07e-1}        & 0.99812     \\ \hline
7               & \multicolumn{1}{l|}{6.88e-2}      & \multicolumn{1}{l|}{3.70e-1}        & \multicolumn{1}{l|}{0.99851}     & \multicolumn{1}{l|}{2.78e-2}      & \multicolumn{1}{l|}{2.21e-1}        & 0.99676     \\ \hline
10              & \multicolumn{1}{l|}{6.13e-2}      & \multicolumn{1}{l|}{3.21e-1}        & \multicolumn{1}{l|}{0.99425}     & \multicolumn{1}{l|}{1.98e-2}      & \multicolumn{1}{l|}{1.83e-1}        & 0.99191     \\ \hline
\end{tabular}
\end{table}

\subsection{Results and Discussion}
Table \ref{tab:NARMA_combined} shows the MSE, normalized mean square error (NRMSE), and Pearson Correlation Coefficient (CC) between $Y$ and $\bar{Y}$ for the square ICE and Pinwheel ICE configurations. The correlation decreases for both the training and testing data as the NARMA order increases for both topologies, with the square configuration yielding marginally better results. On the testing data, MSE and NRMSE increase as the order rises, 
reflecting the increasing complexity in the relationship between the target and input signals. However, both metrics show a slight decrease when moving from NARMA7 to NARMA10. 
This behavior can be attributed to the skewed nature of the loss function, where 85\% of the weight is assigned to the inverted Pearson correlation coefficient, and only 15\% to the MSE. As a result, the best loss value does not necessarily align with the lowest MSE, which explains the non-intuitive decrease in MSE for higher NARMA orders during training. Overall, the imitation performance remains strong as we progress from NARMA2 to NARMA10, with the CC consistently above 99\%.

\begin{figure}
    \centering
    \subfigure[$N=2$]
    {
        \includegraphics[width=0.225\textwidth]{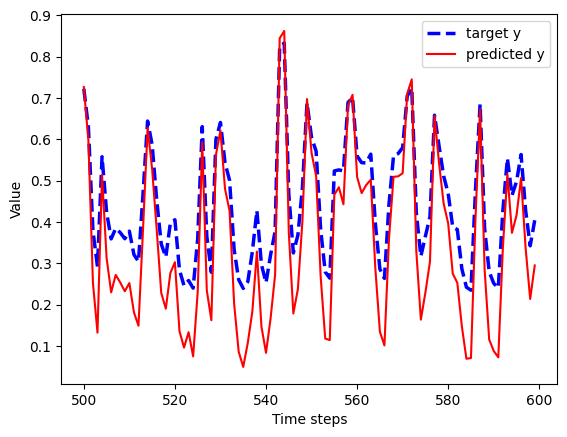}
        \label{fig_narma2:a}
    }
    \subfigure[$N=5$]
    {
        \includegraphics[width=0.225\textwidth]{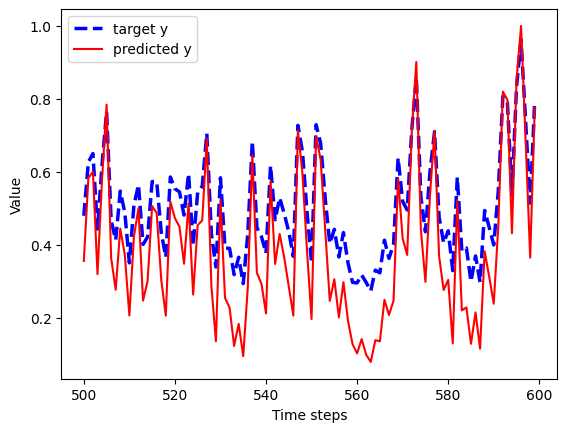}
        \label{fig_narma5:b}
    }
    \subfigure[$N=7$]
    {
        \includegraphics[width=0.225\textwidth]{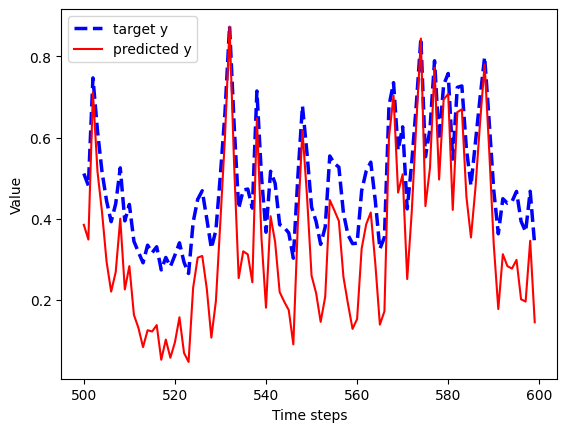}
        \label{fig_narma7:c}
    }
    \subfigure[$N=10$]
    {
        \includegraphics[width=0.225\textwidth]{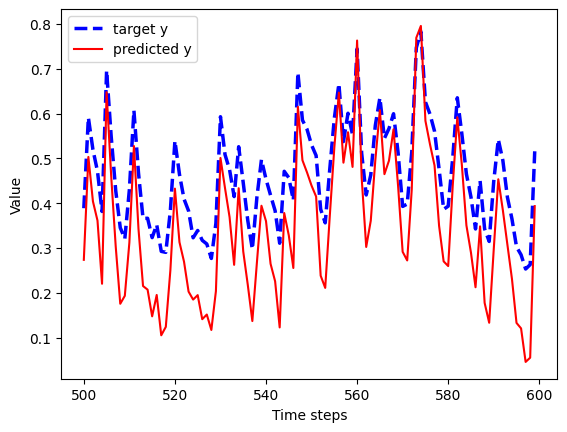}
        \label{fig_narma10:d}
    }
    \caption{NARMA predictions with the square ASI topology, showing improved tracking of the target signal as the order increases from 2 to 10.}
    \label{fig_NarmaVis}
\end{figure}

Figure \ref{fig_NarmaVis} shows a visual comparison between $Y$ and $\bar{Y}$, demonstrating that our method successfully reproduces the NARMA signals across all four orders. By explicitly incorporating past reservoir states, the delay embeddings enhance the reservoir node's memory capacity, allowing for better extraction of dynamical information from the input. This makes it possible to achieve high performance using relatively simple learning architectures such as a multilayer perceptron (MLP).

Table \ref{tab:narma_comp} presents a comparison of the NRMSE results from the proposed method with two other reservoir computing systems \cite{manneschi_exploiting_2021, vatsavai_silicon_2021} on NARMA prediction tasks. For both the NARMA5 and NARMA10 series, our proposed mixed embeddings method, which uses 51 embeddings in the reservoir state vector, outperforms both the unconnected and connected versions of \cite{manneschi_exploiting_2021}, which use 100-dimensional state vectors. This advantage is because \cite{manneschi_exploiting_2021} relies on hierarchical echo state network (ESN) reservoirs, which depend solely on internal recurrence for temporal memory, whereas our method explicitly captures both past and current states using mixed delay/nondelay embeddings.

However, \cite{vatsavai_silicon_2021} slightly outperforms our method on NARMA10. This is due to its use of feedback mechanisms introduced by silicon microrings, which enhance the system's ability to handle long-term memory tasks. On the other hand, their approach involves both input signal masking and encoding, adding complexity to the pre-processing stage.

\begin{table}[H]
\centering
\caption{NRMSE Comparison of the proposed method (using square ice geometry) with two reservoir computing systems for NARMA prediction tasks. $H$ denotes the reservoir state dimensionality.}
\label{tab:narma_comp}
\begin{tabular}{|l|l|l|l|}
\hline
\textbf{Sequence}                 & \textbf{System} & \textbf{H} & \textbf{NRMSE}    \\ \hline
\multirow{4}{*}{\textbf{NARMA10}}   & Connected ESN\cite{manneschi_exploiting_2021}        & 100        & 0.36            \\ \cline{2-4} 
                                & Unconnected ESN\cite{manneschi_exploiting_2021}        & 100        & 0.40         \\ \cline{2-4} 
                                & \textbf{Silicon MR \cite{vatsavai_silicon_2021}}       & 40        & \textbf{0.164}         \\ \cline{2-4} 
                                & Ours   & 51         & 0.169 \\ \hline
\multirow{3}{*}{\textbf{NARMA5}} & Connected ESN\cite{manneschi_exploiting_2021}        & 100         & 0.22         \\ \cline{2-4} 
                                & Unconnected ESN\cite{manneschi_exploiting_2021}        & 100         & 0.30         \\ \cline{2-4} 
                                & \textbf{Ours}   & 51         & \textbf{0.187} \\ \hline
\end{tabular}
\end{table}

\section{Time Series Prediction}
\subsection{Method}

We apply our PNA-based reservoir system to the prediction of the Mackey-Glass time series~\cite{mackey_oscillation_1977}, a common benchmark task in reservoir computing. Different from NARMA, the Mackey-Glass dataset exhibits seasonality, featuring recurring patterns with variations in relation to time. We adopt the parameters of $\beta=0.2$, $\gamma=0.1$, and $n=10$ to generate the dataset $\{u_t\}$. Our experiment is conducted by sweeping the prediction horizon $\tau$, the number of time steps into the future we aim to predict. The specific $\tau$ values listed are $\{1,4,8,12,16,20,24,28,32,48,64\}$, categorized as short-term predictions for $\tau\leq 15$ and long-term predictions otherwise.

Following the generation of mixed embeddings from $0\sim1$ scaled $\{u_t\}$ as shown in Fig.~\ref{fig_proposed:1a} with $h=50$, we fully train the perceptron with output $\Bar{Y}=\{\Bar{y}_t\mid\Bar{y}_t=f(\{\mathbf{r}\}_t)\}$, as shown in Fig.~\ref{fig_proposed:1b}. Our Mackey-Glass dataset contains $3000$ samples, with a train/test ratio of $2:1$.

\subsection{Short-Term Prediction: Results and Discussion}

\begin{table}[]
\centering
\caption{Short-term Mackey-Glass prediction results for different horizons ($\tau$) using two $7\times 7$ topologies.}
\label{tab:mg_short}
\begin{tabular}{|l|llllll|}
\hline
$\bm{\tau}$ & \multicolumn{1}{l|}{\textbf{MSE}} & \multicolumn{1}{l|}{\textbf{NRMSE}} & \multicolumn{1}{l|}{\textbf{CC}} & \multicolumn{1}{l|}{\textbf{MSE}} & \multicolumn{1}{l|}{\textbf{NRMSE}} & \textbf{CC} \\ \hline
\textbf{}       & \multicolumn{3}{c|}{\textbf{Train (Square ICE)}}                                                                        & \multicolumn{3}{c|}{\textbf{Test (Square ICE)}}                                                    \\ \hline
1               & \multicolumn{1}{l|}{1.82e-4}      & \multicolumn{1}{l|}{1.35e-2}        & \multicolumn{1}{l|}{0.99955}     & \multicolumn{1}{l|}{1.77e-4}      & \multicolumn{1}{l|}{1.33e-2}        & 0.99956     \\ \hline
4               & \multicolumn{1}{l|}{5.65e-5}      & \multicolumn{1}{l|}{7.52e-3}        & \multicolumn{1}{l|}{0.99965}     & \multicolumn{1}{l|}{5.96e-5}      & \multicolumn{1}{l|}{7.73e-3}        & 0.99965     \\ \hline
8               & \multicolumn{1}{l|}{2.18e-4}      & \multicolumn{1}{l|}{1.48e-2}        & \multicolumn{1}{l|}{0.99937}     & \multicolumn{1}{l|}{2.38e-4}      & \multicolumn{1}{l|}{1.54e-2}        & 0.99935     \\ \hline
12              & \multicolumn{1}{l|}{2.35e-4}      & \multicolumn{1}{l|}{1.53e-2}        & \multicolumn{1}{l|}{0.99952}     & \multicolumn{1}{l|}{2.45e-4}      & \multicolumn{1}{l|}{1.57e-2}        & 0.99951     \\ \hline
\textbf{}       & \multicolumn{3}{c|}{\textbf{Train (Pinwheel ICE)}}                                                                        & \multicolumn{3}{c|}{\textbf{Test (Pinwheel ICE)}}                                                    \\ \hline
1                               & \multicolumn{1}{l|}{1.35e-4}      & \multicolumn{1}{l|}{1.16e-2}        & \multicolumn{1}{l|}{0.99959}     & \multicolumn{1}{l|}{1.32e-4}      & \multicolumn{1}{l|}{1.15e-2}        & 0.99959     \\ \hline
4                               & \multicolumn{1}{l|}{2.46e-5}      & \multicolumn{1}{l|}{4.96e-3}        & \multicolumn{1}{l|}{0.99982}     & \multicolumn{1}{l|}{2.79e-5}      & \multicolumn{1}{l|}{5.29e-3}        & 0.99981     \\ \hline
8                               & \multicolumn{1}{l|}{2.06e-4}      & \multicolumn{1}{l|}{1.44e-2}        & \multicolumn{1}{l|}{0.99914}     & \multicolumn{1}{l|}{1.98e-4}      & \multicolumn{1}{l|}{1.41e-2}        & 0.99908     \\ \hline
12                              & \multicolumn{1}{l|}{5.38e-4}      & \multicolumn{1}{l|}{2.32e-2}        & \multicolumn{1}{l|}{0.99939}     & \multicolumn{1}{l|}{7.56e-4}      & \multicolumn{1}{l|}{2.75e-2}        & 0.99902     \\ \hline
\end{tabular}
\end{table}

\begin{table}[]
\centering
\caption{Comparisons of short-term ($\tau=1$) Mackey-Glass prediction results. Our best result is listed here. $H$ denotes the reservoir state dimensionality.}
\label{tab:mg_short_comp}
\begin{tabular}{|l|l|l|l|l|}
\hline
\textbf{Metric}                 & \textbf{System} & \textbf{H} & \textbf{Train}   & \textbf{Test}    \\ \hline
\multirow{2}{*}{MSE}   & ASVI~\cite{gartside_reconfigurable_2022-2}        & 397        & 1.69e-3          & 2.75e-3          \\ \cline{2-5} 
                                & \textbf{Ours}   & 51         & \textbf{1.35e-4} & \textbf{1.32e-4} \\ \hline
\multirow{2}{*}{NRMSE} & \cite{hu_distinguishing_2023}        & 36         & 1.72e-1          & 1.76e-1          \\ \cline{2-5} 
                                & \textbf{Ours}   & 51         & \textbf{1.16e-2} & \textbf{1.15e-2} \\ \hline
\end{tabular}
\end{table}

\begin{figure}
    \centering
    \includegraphics[scale=0.5]{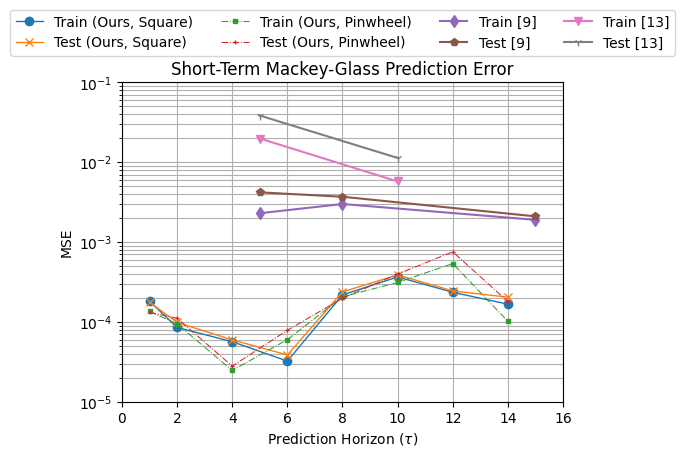}
    \caption{Comparison of short-term Mackey-Glass prediction error against implementations in prior works. Our system outperforms designs with much higher reservoir state dimensionality, by an order of magnitude or more.}
    \label{fig_mg_short_comp}
\end{figure}

\begin{figure}
    \centering
    \includegraphics[scale=0.5]{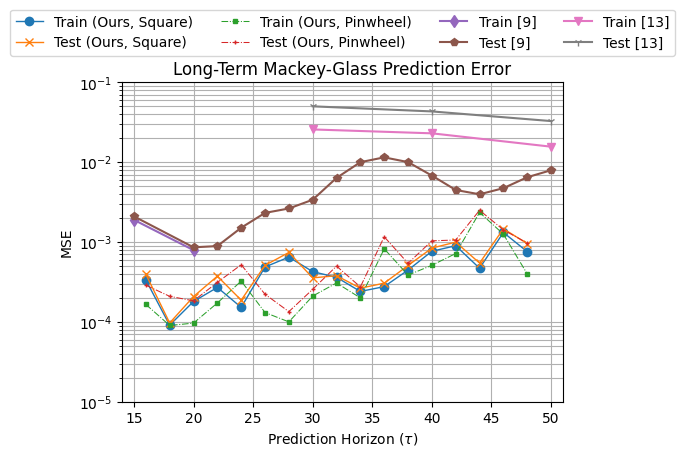}
    \caption{Comparison of long-term Mackey-Glass prediction error. Our system continues to significantly outperform high-dimensional designs in prior works. }
    \label{fig_mg_long_comp}
\end{figure}

\begin{table}[]
\centering
\caption{Short-term Mackey-Glass prediction ($\tau=8$) with a single readout layer, using two $7\times7$ topologies. With an increasing history length $h$, the performance of the single layer implementation approaches that of the multi-layer perceptron.}
\label{tab:mg_short_single_layer}
\begin{tabular}{|l|l|l|}
\hline
$\bm{h}$ & \textbf{CC (Square)} & \textbf{CC (Pinwheel)} \\ \hline
50         & 0.91999              & 0.87257                \\ \hline
100        & 0.95527              & 0.95527                \\ \hline
150        & 0.97917              & 0.98162                \\ \hline
200        & 0.99955              & 0.99935                \\ \hline
\end{tabular}
\end{table}

As shown in TABLE~\ref{tab:mg_short}, for smaller $\tau$ values, our single-node delay/nondelay-embedding PNA system consistently achieves high accuracy, obtaining CC values above $0.999$ and MSE below $0.001$ for all short-term prediction scenarios. TABLE~\ref{tab:mg_short_comp} shows that at the shortest-term ($\tau=1$) prediction task, our system outperforms previous reservoir designs in MSE and NRMSE by more than an order of magnitude. In other prediction scenarios illustrated in Fig.~\ref{fig_mg_short_comp}, our errors are also well below those of the competitions. Notably, our system achieves this performance advantage with a (much) smaller architecture. Some previous systems adopt high numbers of embeddings in the reservoir state vector (e.g., $1601$ reservoir outputs per single field input in \cite{lee_task-adaptive_2024}, and $397$ reservoir outputs per input in \cite{gartside_reconfigurable_2022-2}). Our system uses a single reservoir node with 51 embeddings.

Our system's high performance is attributable to our unique design of mixed delay/nondelay embeddings, which enhances the memory capacity by explicitly incorporating past reservoir states. This design makes the dynamical information in the input signal more easily extracted by the relatively simple readout network. This observation is supported by the experiment that predicts for $\tau=8$ (shown in TABLE~\ref{tab:mg_short_single_layer}), where we used a single linear layer in place of the perceptron ($f(\cdot)$ in Fig.~\ref{fig_proposed:1b}). As the history length $h$ increases, the prediction performance of the single layer design catches up with the original design with the perceptron. This experiment reveals a tradeoff between the reservoir state dimensionality, $H$, and the complexity of the readout network.

\begin{table}[]
\centering
\caption{Long-term Mackey-Glass prediction results for the PNA with Square geometric topology.}
\label{tab:mg_long}
\begin{tabular}{|l|llllll|}
\hline
$\bm{\tau}$ & \multicolumn{1}{l|}{\textbf{MSE}} & \multicolumn{1}{l|}{\textbf{NRMSE}} & \multicolumn{1}{l|}{\textbf{CC}} & \multicolumn{1}{l|}{\textbf{MSE}} & \multicolumn{1}{l|}{\textbf{NRMSE}} & \textbf{CC} \\ \hline
\textbf{}       & \multicolumn{3}{c|}{\textbf{Train (Square ICE)}}                                                                        & \multicolumn{3}{c|}{\textbf{Test (Square ICE)}}                                                    \\ \hline
16              & \multicolumn{1}{l|}{3.36e-4}      & \multicolumn{1}{l|}{1.83e-2}        & \multicolumn{1}{l|}{0.99939}     & \multicolumn{1}{l|}{3.96e-4}      & \multicolumn{1}{l|}{1.99e-2}        & 0.99937     \\ \hline
20              & \multicolumn{1}{l|}{1.80e-4}      & \multicolumn{1}{l|}{1.34e-2}        & \multicolumn{1}{l|}{0.99963}     & \multicolumn{1}{l|}{2.08e-4}      & \multicolumn{1}{l|}{1.44e-2}        & 0.99962     \\ \hline
24              & \multicolumn{1}{l|}{1.52e-4}      & \multicolumn{1}{l|}{1.23e-2}        & \multicolumn{1}{l|}{0.99894}     & \multicolumn{1}{l|}{1.87e-4}      & \multicolumn{1}{l|}{1.37e-2}        & 0.99888     \\ \hline
28              & \multicolumn{1}{l|}{6.48e-4}      & \multicolumn{1}{l|}{2.55e-2}        & \multicolumn{1}{l|}{0.99784}     & \multicolumn{1}{l|}{7.42e-4}      & \multicolumn{1}{l|}{2.73e-2}        & 0.99769     \\ \hline
32              & \multicolumn{1}{l|}{3.62e-4}      & \multicolumn{1}{l|}{1.90e-2}        & \multicolumn{1}{l|}{0.99799}     & \multicolumn{1}{l|}{3.79e-4}      & \multicolumn{1}{l|}{1.95e-2}        & 0.99789     \\ \hline
48              & \multicolumn{1}{l|}{7.58e-4}      & \multicolumn{1}{l|}{2.75e-2}        & \multicolumn{1}{l|}{0.99790}     & \multicolumn{1}{l|}{9.58e-4}      & \multicolumn{1}{l|}{3.10e-2}        & 0.99759     \\ \hline
64              & \multicolumn{1}{l|}{8.15e-3}      & \multicolumn{1}{l|}{9.03e-2}        & \multicolumn{1}{l|}{0.94450}     & \multicolumn{1}{l|}{8.26e-3}      & \multicolumn{1}{l|}{9.10e-2}        & 0.94301     \\ \hline
\textbf{}       & \multicolumn{3}{c|}{\textbf{Train (Pinwheel ICE)}}                                                                        & \multicolumn{3}{c|}{\textbf{Test (Pinwheel ICE)}}                                                    \\ \hline
16              & \multicolumn{1}{l|}{1.66e-4}      & \multicolumn{1}{l|}{1.29e-2}        & \multicolumn{1}{l|}{0.99905}     & \multicolumn{1}{l|}{2.88e-4}      & \multicolumn{1}{l|}{1.70e-2}        & 0.99827     \\ \hline
20              & \multicolumn{1}{l|}{9.66e-5}      & \multicolumn{1}{l|}{9.83e-3}        & \multicolumn{1}{l|}{0.99973}     & \multicolumn{1}{l|}{1.84e-4}      & \multicolumn{1}{l|}{1.36e-2}        & 0.99938     \\ \hline
24              & \multicolumn{1}{l|}{3.25e-4}      & \multicolumn{1}{l|}{1.80e-2}        & \multicolumn{1}{l|}{0.99809}     & \multicolumn{1}{l|}{5.20e-4}      & \multicolumn{1}{l|}{2.28e-2}        & 0.99763     \\ \hline
28              & \multicolumn{1}{l|}{1.00e-4}      & \multicolumn{1}{l|}{1.00e-2}        & \multicolumn{1}{l|}{0.99945}     & \multicolumn{1}{l|}{1.35e-4}      & \multicolumn{1}{l|}{1.16e-2}        & 0.99916     \\ \hline
32              & \multicolumn{1}{l|}{3.08e-4}      & \multicolumn{1}{l|}{1.75e-2}        & \multicolumn{1}{l|}{0.99808}     & \multicolumn{1}{l|}{5.00e-4}      & \multicolumn{1}{l|}{2.24e-2}        & 0.99746     \\ \hline
48              & \multicolumn{1}{l|}{3.99e-4}      & \multicolumn{1}{l|}{2.00e-2}        & \multicolumn{1}{l|}{0.99835}     & \multicolumn{1}{l|}{9.73e-4}      & \multicolumn{1}{l|}{3.12e-2}        & 0.99644     \\ \hline
64              & \multicolumn{1}{l|}{8.46e-3}      & \multicolumn{1}{l|}{9.20e-2}        & \multicolumn{1}{l|}{0.94315}     & \multicolumn{1}{l|}{8.68e-3}      & \multicolumn{1}{l|}{9.33e-2}        & 0.94128     \\ \hline
\end{tabular}
\end{table}

\begin{figure}
    \centering
    \subfigure[$\tau=6$]
    {
        \includegraphics[width=0.225\textwidth]{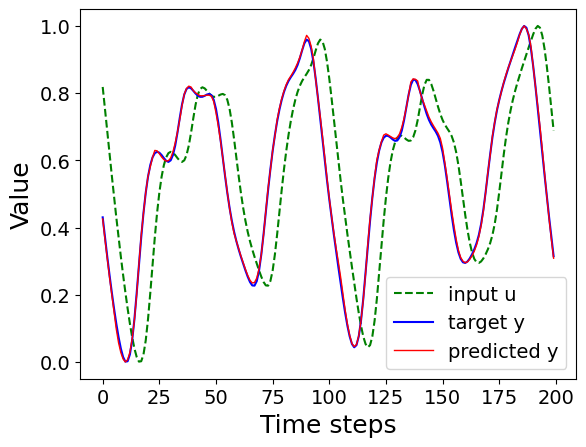}
        \label{fig_ex:a}
    }
    \subfigure[$\tau=12$]
    {
        \includegraphics[width=0.225\textwidth]{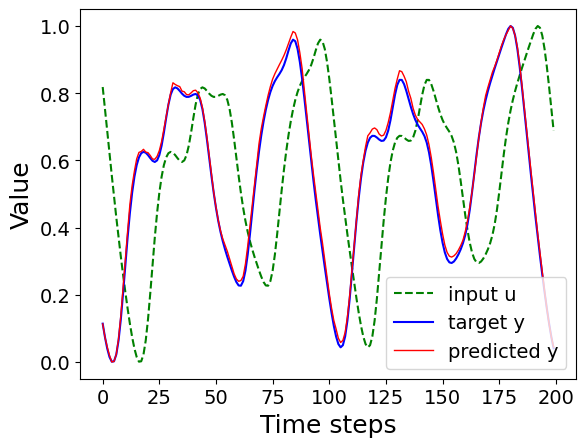}
        \label{fig_ex:b}
    }
    \subfigure[$\tau=24$]
    {
        \includegraphics[width=0.225\textwidth]{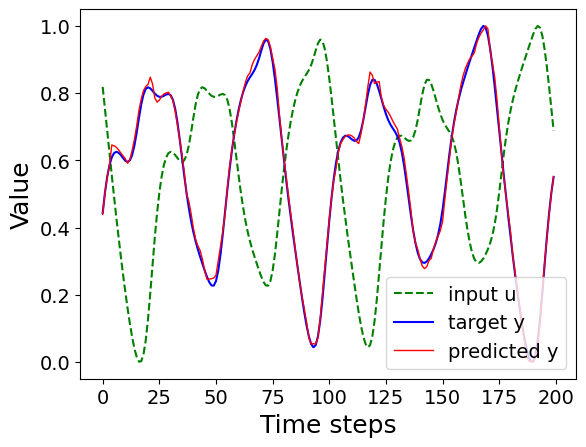}
        \label{fig_ex:c}
    }
    \subfigure[$\tau=40$]
    {
        \includegraphics[width=0.225\textwidth]{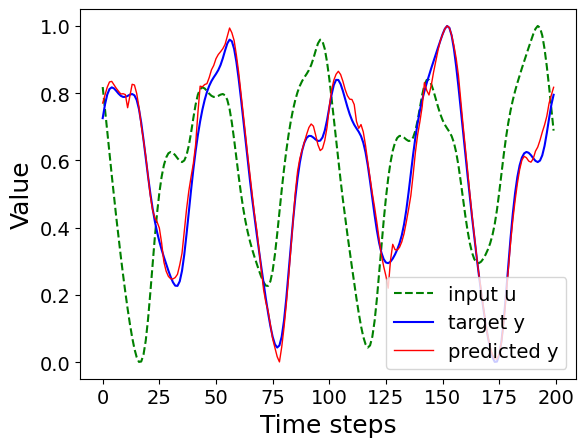}
        \label{fig_ex:d}
    }
    \caption{Illustration of Mackey-Glass predictions with the square ASI topology, for different horizon ($\tau$) values. As $\tau$ increases, the high-frequency sections in the signal become more error-prone.}
    \label{fig_ex:1}
\end{figure}

\subsection{Long-Term Prediction: Results and Discussion}


First, it is important to note that no prior work has investigated predictions for $\tau>15$ in the main body of the publication, although \cite{lee_task-adaptive_2024} and \cite{gartside_reconfigurable_2022-2} have explored a higher $\tau$ in their supplemental work. Under our default history length $h=50$, we sweep for $\tau$ up to 48, as shown in TABLE \ref{tab:mg_long}. Fig.~\ref{fig_mg_long_comp} shows our system outperforms others by a wide margin, albeit having the smallest reservoir state dimensionality. Within our system, we observed that the square reservoir topology performs this prediction task better, as evidenced by the narrower gap between training and testing errors than with the pinwheel topology. Figs.~\ref{fig_ex:c} and \ref{fig_ex:d} showcase two long-term predicted signals, where an increasing amount of oscillating noise appears in high-frequency ranges of the signal. Finally, $\tau=64$ is attempted, with significantly deteriorated error metrics compared with $\tau<50$, highlighting the need to further increase the history length for longer-term predictions.

\section{Reservoir Quality and Overheads}
To explain the
improved
performance
of
our reservoir system, we evaluated
key
reservoir quality metrics such as memory capacity (MC) and nonlinearity (NL),
as defined in
\cite{love_task_2021}.
Using the past $k=8$ input signal samples and non-delayed embeddings, our PNA reservoir system yielded $\text{MC}=5.71$ and $\text{NL}=0.81$. 
These values are higher than the spectral fingerprinting-based reservoir system from \cite{stenning_neuromorphic_2023} (with a single, square-geometry PNA reservoir node) that has $\text{MC}=2.6$ and $\text{NL}=0.38$. The higher MC and NL for our PNA reservoir naturally
contributed to
better performance for the time-series imitation task. However, $\text{MC}=5.71$ for $k=8$ means that the reservoir can memorize only the past $5.71$ input samples
out of the past
$k=8$ samples, which limits its suitability for long-term time-series prediction.
Increasing the input sample sizes to $k=100$ yielded only $\text{MC}=35.34$ and $\text{NL}=0.15$, which did not lead to good prediction performance. On the other hand, the incorporation of our proposed delay embeddings significantly raised MC and NL to $7.99$ and $0.99$ for $k=8$  and $90.26$ and $0.99$ for $k=100$.
The improved MC and NL values increased the performance of our PNA reservoir system for the imitation and prediction tasks compared to the other PNA reservoir systems from prior work.  

It is worth noting that this increase in the performance of our reservoir system does not
come at an increased overhead of storing the reservoir state embeddings. For instance, the spectral fingerprinting based PNA reservoir system from \cite{lee_task-adaptive_2024} requires a total of $1601$ reservoir state embeddings (each embedding is a magnitude of a ferromagnetic resonance frequency) with 16-bit storage overhead per embedding, making the total storage overhead of $25{\small,}616$ bits. In contrast, our reservoir system requires $51$ embeddings ($1$ nondelay and $50$ delay embeddings) with each embedding containing a total of $112$ magnetization states of up to $3$-bit each ($3$-bit because up to eight magnetization levels can be probed using GMR-based readout \cite{hu_distinguishing_2023}). Therefore, the total overhead for storing the mixed delay/nondelay embeddings of our reservoir system is $17{\small,}136$ bits. Thus, our delay/nondelay embeddings-based PNA reservoir system provides better computing performance at a lower storage overhead compared to the other PNA reservoirs from prior work.

\section{Conclusions and Future Work}
We presented a TMR probing-based PNA reservoir system. Our system employs a single PNA reservoir node that combines the delay and nondelay spatial embeddings of the reservoir state information at the output layer. The combination of delay- and nondelay-based spatial embeddings dramatically increased the richness of the spatially separable dynamical information readily available to our reservoir system, increasing the imitation and prediction accuracy of our system compared to the TMR/GMR/AMR-based PNA reservoir systems from prior works. Based on our analysis, our reservoir system outperforms existing PNA-based reservoir systems for the imitation of NARMA 2, NARMA 5, NARMA 7, and NARMA 10 time series data, and for the short-term and long-term prediction of the Mackey Glass time series data. Our results corroborate the excellent capabilities of our PNA-based reservoir computing system for practical, efficient, and accurate processing of time series data. 

In the future, we will look to extend this work in the following four ways. First, we will compare the time-series imitation and prediction accuracy of our system with other reservoir-less neural network architectures (e.g., 1D convolutional neural network) as well as other software reservoir systems (e.g., echo state network). Second, we will perform a sensitivity analysis to investigate the lowest possible value of h (the required number of delay embeddings) that is achievable without notably compromising the computing accuracy. Third, we will investigate the use of our reservoir system for additional single variate and multivariate time-series datasets. Fourth, we will explore the physical hyperparameters of PNAs to investigate their impacts on various reservoir quality metrics such as MC, NL, stability, echo state property, and expressivity.



\section*{Acknowledgment}
This material is based upon work supported by the U.S. Department of Energy, Office of Science, Office of Basic Energy Sciences under Award Number DE-SC-0024346. This research used resources of the Advanced Light Source, a U.S. DOE Office of Science User Facility under contract no. DE-AC02-05CH11231.




\bibliographystyle{IEEEtran}
\bibliography{Thakkar-Bib-Library, references}

\end{document}